\documentclass[twocolumn]{webofc}
\usepackage[varg]{txfonts}   % Web of Conferences font

\newcommand\beq{\begin{equation}}
\newcommand\eeq{\end{equation}}
\newcommand\beqa{\begin{eqnarray}}
\newcommand\eeqa{\end{eqnarray}}

\newcommand{\al}{\alpha}

\begin{document}
\title{Stability of the homogeneous steady state for a model of a confined quasi-two-dimensional granular fluid}

\author{\firstname{Vicente} \lastname{Garz\'o}\inst{1}\fnsep\thanks{\email{vicenteg@unex.es}} \and
        \firstname{Ricardo} \lastname{Brito}\inst{2}
        %\fnsep\thanks{\email{brito@ucm.es}}
             \and
        \firstname{Rodrigo} \lastname{Soto}\inst{3}
       %\fnsep\thanks{\email{rsoto@dfi.uchile.cl}}
        % etc.
}

\institute{Departamento de
F\'{\i}sica and Instituto de Computaci\'on Cient\'{\i}fica Avanzada (ICCAEx), Universidad de Extremadura, E-06071 Badajoz, Spain
\and
           Departamento de Estructura de la Materia, F\'{\i}sica T\'ermica y Electr\'onica and GISC, Universidad Complutense de Madrid, Spain
\and
           Departamento de
F\'{\i}sica, Facultad de Ciencias F\'{\i}sicas y Matem\'aticas, Universidad de Chile, Santiago, Chile
          }

\abstract{
A linear stability analysis of the hydrodynamic equations of a model for confined quasi-two-dimensional granular gases is carried out. The stability analysis is performed around the homogeneous steady state (HSS) reached eventually by the system after a transient regime.
In contrast to previous studies (which considered dilute or quasielastic systems), our analysis is based on the results obtained from the \emph{inelastic} Enskog kinetic equation, which takes into account the (nonlinear) dependence of the transport coefficients and the cooling rate on dissipation and applies to moderate densities. As in earlier studies, the analysis shows that the HSS is linearly stable with respect to long enough wavelength excitations.
}
\maketitle

\section{Introduction}
\label{sec1}

It is well established that granular matter can achieve a rapid flow regime when grains are subjected to a violent and sustained excitation. Under these conditions, the motion of grains is quite similar to that of a gas of activated collisional grains and hence, they can be modeled as a gas of smooth inelastic hard spheres \cite{BP04,G19}. Since the kinetic energy is lost by collisions one has to inject energy into the system to maintain it under rapid flow conditions. There are several ways in real experiments of supplying energy to the system; for instance, by shearing or vibrating its walls \cite{YHCMW02} or alternatively by bulk driving \cite{SGS05,AD06}. However, this type of heating produces in general strong spatial gradients in the bulk region so that, a theoretical description that goes beyond the conventional Navier-Stokes (NS) description (which applies for small spatial gradients) is required to offer a complete hydrodynamic description. Therefore, due to the intricacies associated with the study of the above situations, it is quite common in theoretical and computational studies to introduce external driving forces (thermostats) that supply energy to compensate for the collisional cooling and so, the gas reaches a stationary nonequilibrium state \cite{G19}.

An alternative to the use of external forces has been proposed in the past few years \cite{MS16}. The idea is to employ a particular geometry (quasi-two-dimensional geometry) where the granular gas is confined in a box that is vertically vibrated and hence, energy is injected into the vertical degrees of freedom of particles via the collisions of grains with the top and bottom plates. The energy gained by collisions with the walls is then transferred to the horizontal degrees of freedom by collisions between grains. Under these conditions, when the system is observed from above, it is fluidized and can remain homogeneous.  A collisional model for the transfer of energy from the vertical to horizontal degrees of freedom in the quasi-two-dimensional geometry was proposed a few years ago \cite{BRS13}. In this model, an extra velocity $\Delta$ is added to the relative motion of colliding spheres so that, the magnitude of the normal component of the relative velocity of colliding spheres is increased by a given factor in the collision. This term mimics the transfer of energy from the vertical degrees of freedom to the horizontal ones. The $\Delta$-model has been widely employed in the past few years to derive the NS hydrodynamic equations for monocomponent \emph{dilute} \cite{BBMG15} and \emph{dense} \cite{GBS18} granular gases with explicit expressions of the transport coefficients. Some recent works \cite{BSG20} have extended previous efforts to
%the case of
multicomponent granular gases.

The knowledge of the NS transport coefficients and the cooling rate for monocomponent granular gases opens up the possibility of analyzing the stability of the so-called homogeneous steady state (HSS). This is a quite relevant state of confined quasi-two-dimensional systems. To the best of our knowledge, two different papers have studied the stability of HSS. For \emph{dilute} granular gases, a linear stability analysis of the hydrodynamic equations was performed in Ref.\ \cite{BBGM16}. For that, the expressions obtained in Ref.\ \cite{BBMG15} of the NS transport coefficients were used; these expressions take into account the nonlinear dependence of transport coefficients on the coefficient of restitution $\alpha$. The stability analysis was extended to moderate dense gases \cite{BRS13} but considering the \emph{elastic} forms of the NS transport coefficients. Both works conclude that the HSS is linearly stable. On the other hand, although the predictions of both works \cite{BBMG15,BRS13} compares well with computer simulations, it is worth to assess to whether, and if so to what extent, the results obtained before \cite{BBMG15,BRS13} may be altered when the improved forms of the \emph{inelastic} transport coefficients of a \emph{moderate dense} granular gas \cite{GBS18} are considered. This is the main goal of the present contribution.

\section{Hydrodynamic equations}
\label{sec2}

We consider a granular fluid composed of smooth inelastic hard spheres of mass $m$ and diameter $\sigma$. Collisions are
characterized by a positive (constant) coefficient of normal restitution $\alpha\leq 1$, such the normal component of the relative velocity $g_\text{n}$ changes to $g_{\text{n}}'=-\alpha g_\text{n} - 2\Delta$. Here, $g_{\text{n}}'$ is the post-collisional normal component of the relative velocity and $\Delta>0$ is an extra velocity added to $g_\text{n}$. At a kinetic level, all the relevant information on the system is given through the one-particle velocity distribution function, which is assumed to obey the (inelastic) Enskog equation. The corresponding
(macroscopic) hydrodynamic equations for the number density $n({\bf r}, t)$, the flow velocity ${\bf U}({\bf r}, t)$, and the local temperature $T({\bf r}, t)$ can be easily derived from the Enskog kinetic equation. In the context of the so-called $\Delta$-model, the hydrodynamic equations are \cite{BBMG15,GBS18}
\begin{equation}
D_{t}n+n \nabla \cdot {\bf U}=0\;,  \label{2.1}
\end{equation}
\begin{equation}
\rho D_{t}U_i+\partial_j P_{ij}=0\;,  \label{2.2}
\end{equation}
\begin{equation}
D_{t}T+\frac{2}{d n}\left(\partial_i q_i+P_{ij}\partial_j U_i\right)=-\zeta T.
\label{2.3}
\end{equation}
In the above equations, $d$ is the dimensionality of the system, $D_{t}=\partial _{t}+{\bf U}\cdot \nabla $
is the material derivative,  $\rho=nm$ is the mass density, $\mathsf{P}$ is the pressure tensor, ${\bf q}$ is the heat flux, and $\zeta$
is the cooling rate due to the energy dissipated in collisions. Equations \eqref{2.1}--\eqref{2.3} become a closed set of differential equations for the hydrodynamic fields once the fluxes and the cooling rate are expressed in terms of them. The detailed form of
the constitutive equations and the transport coefficients appearing in them have been derived in Ref.\ \cite{GBS18} in the context of the (inelastic) Enskog equation. To first order in the spatial gradients (NS hydrodynamic order), the corresponding constitutive equations are
\begin{equation}
P_{ij}=p\delta _{ij}-\eta \left( \partial_{j}U_{i }+\partial_{i
}U_{j}-\frac{2}{d}\delta_{ij} \nabla \cdot \mathbf{U}\right)-\gamma \delta_{ij} \nabla \cdot {\bf U},
\label{2.4}
\end{equation}
\begin{equation}
{\bf q}=-\kappa \nabla T-\mu \nabla n,
\label{2.5}
\end{equation}
\begin{equation}
\label{2.6}
\zeta=\zeta^{(0)}+\zeta_U \nabla \cdot {\bf U},
\end{equation}
where $p$ is the hydrostatic pressure, $\eta$ is the shear
viscosity, $\gamma$ is the bulk viscosity, $\kappa$ is the thermal
conductivity, and $\mu$ is a new transport coefficient not present
in the elastic case. For general time-dependent states, the expressions for the pressure, the transport
coefficients and the cooling rate can be written, respectively, in the forms $p=nT p^*(\alpha, \phi,\Delta^*)$, $\eta=\eta_0 \eta^*(\alpha, \phi,\Delta^*)$, $\gamma=\eta_0 \gamma^*(\alpha, \phi,\Delta^*)$, $\kappa=\kappa_0 \kappa^*(\alpha, \phi,\Delta^*)$, $\mu=(T\kappa_0/n) \mu^*(\alpha, \phi,\Delta^*)$, and $\zeta^{(0)}=(nT/\eta_0) \zeta_0^*(\alpha, \phi,\Delta^*)$. Here, $\phi$ is the solid volume fraction, $\eta_0=(d+2/8)\Gamma\Big(d/2\Big)\pi^{-\frac{d-1}{2}}\sigma^{1-d}\sqrt{mT}$ and $\kappa_0=[d(d+2)/(2(d-1))](\eta_0/m)$ are the low-density values of the shear viscosity and thermal conductivity, respectively, for elastic collisions. In addition, the (reduced) velocity is $\Delta^*=\Delta/v_\text{th}$, $v_\text{th}=\sqrt{2T/m}$ being the thermal velocity.
\begin{figure}[h]
\centering
\includegraphics[width=0.35 \textwidth,clip]{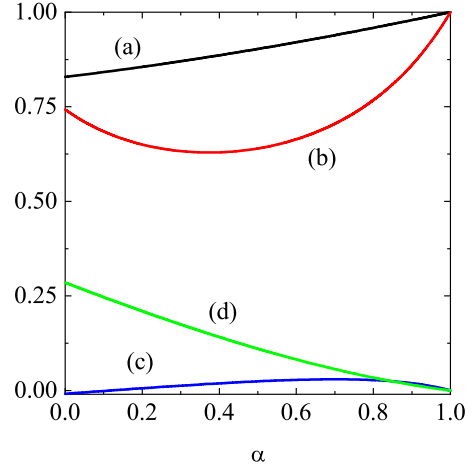}
\caption{Plot of $\eta^*(\al,\phi)/\eta^*(1,\phi)$ (a), $\kappa^*(\al,\phi)/\kappa^*(1,\phi)$ (b), $\mu^*(\al,\phi)=n|\mu(\al,\phi)|/(T\kappa_0)$ (c), and $|\zeta_U|$ (d) versus the coefficient
of restitution $\al$ for $d=2$ and $\phi=0.2$.}
\label{fig1}       % Give a unique label
\end{figure}

As mentioned in previous papers \cite{BBMG15,GBS18}, it is quite apparent that the dependence of the transport coefficients on the temperature in the $\Delta$-model is in general much more intricate than in the conventional inelastic hard sphere (IHS) model \cite{G19}. This is due essentially to the dependence of the scaled coefficients $p^*$, $\eta^*$, $\gamma^*$, $\kappa^*$, $\mu^*$, $\zeta_0^*$, and $\zeta_U$ on the (dimensionless) velocity $\Delta^*$. However, a simple but interesting situation corresponds to the HSS where the granular temperature achieves a constant value in the long-time limit. The steady value of temperature is determined from the condition $\zeta_0^*=0$ and so, $\Delta_\text{s}^*$ is a function of  $\al$ given by
\beq
\label{2.11}
\Delta_\text{s}^*(\al)=\frac{1}{2}\sqrt{\frac{\pi}{2}}\al \Bigg[\sqrt{1+\frac{4(1-\al^2)}{\pi \al^2}}-1\Bigg].
\eeq
The scaled transport coefficients $p^*$, $\eta^*$, $\gamma^*$, $\kappa^*$, and $\mu^*$ have been recently obtained as functions of both $\al$ and $\phi$ in the HSS. Their forms for $d=2$ can be found in Table I of Ref.\ \cite{GBS18}.

The expression of $\zeta_U$ can be also derived by following similar steps as those made in the IHS model \cite{GD99}. After a simple algebra, one gets
\beq
\label{2.12}
\zeta_U=\frac{2^{d-1}}{d}\phi \chi \Bigg[\Delta_\text{s}^{*2}+\frac{2^{5/2}}{\sqrt{\pi}}\al \Delta_\text{s}^*-\frac{3}{2}(1-\al^2)\Bigg],
\eeq
where $\chi(\phi)$ is the pair correlation function at contact. Note that $\zeta_U=0$ for both elastic collisions ($\al=1$) and/or dilute systems ($\phi=0$) \cite{note}.

Figure \ref{fig1} shows the dependence of the scaled NS transport coefficients and the first-order contribution $\zeta_U$ to the cooling rate on $\al$ for a two-dimensional ($d=2$) confined system at $\phi=0.2$. We observe first that the influence of dissipation on the thermal conductivity is more significant than the one found for the shear viscosity. Moreover, although not shown here, the coefficient $\mu^*$ is always negative for moderate densities ($\phi\lesssim 0.3$ for $d=2$) and its magnitude is tiny for any density. On the other hand, the magnitude $|\zeta_U|$ is larger than that of $|\mu^*|$ and increases with increasing inelasticity.

When the expressions of the pressure tensor, the heat flux and the cooling rate are substituted
into the balance equations \eqref{2.1}--\eqref{2.3} one gets the corresponding NS (closed)
hydrodynamic equations for $n$, ${\bf U}$ and $T$. As has been widely noted in some previous papers \cite{BBMG15,G05}, terms up to second order in the gradients in the expression (\ref{2.6}) for the cooling rate $\zeta$ should be considered in the NS hydrodynamic equation for the granular temperature. This is because these terms are of the same order than the terms coming from the pressure tensor and the heat flux in the above hydrodynamic equation. However, it has been shown \cite{BDKS98} in the conventional IHS model for low-density gases that the contributions from the cooling rate of second order are negligible as compared with the corresponding contributions from Eqs.\ (\ref{2.4})--(\ref{2.6}). It is assumed here that the same holds in the dense case.

\section{Stability analysis}
\label{sec3}

It is quite apparent that the NS hydrodynamic equations admit the existence of a HSS, namely, a uniform state ($\mathbf{U}_\text{H}=\textbf{0}$ without loss of generality) where the steady temperature $T_\text{H}$ is determined from the equation $\zeta^{(0)}(n_\text{H},T_\text{H})=0$. Here, the subscripts H denotes the homogeneous steady state. This state has been widely studied in several previous papers \cite{BRS13,SDB14,BGMB13} and the theoretical results compare quite well with computer simulations. Our aim here is to analyze the stability of the HSS, namely, to investigate if the HSS is stable or unstable with respect to long enough wavelength perturbations. To provide an answer to the above question, it is convenient to perform a \emph{linear} stability analysis of the nonlinear NS hydrodynamic equations with respect to the HSS for \emph{small} initial perturbations.

We assume that the deviations $\delta y_{\alpha}({\bf r},t)=y_{\alpha}({\bf r},t)-y_{\text{H} \alpha}$ are small, where,
$\delta y_{\alpha}({\bf r},t)$ denotes the deviation of $\{n, {\bf U}, T,\}$ from their values in the HSS. To compare with the results derived years ago in the IHS model \cite{G05}, we consider the same time and space variables: $\tau=\frac{1}{2}\nu_{\text{H}}t$ and ${\boldsymbol{\ell}}=\frac{1}{2}(\nu_{\text{H}}/v_{0\text{H}})\mathbf{r}$, where $\nu_{\text{H}}=n_\text{H}T_{\text{H}}/\eta_{0\text{H}}$ and $v_{0\text{H}}=\sqrt{T_{\text{H}}/m}$. The dimensionless time scale $\tau$ is is a measure of the average number of collisions per particle
in the time interval between $0$ and $t$. The unit length $v_{0,\text{H}}/\nu_{\text{H}}$ is proportional to the time-independent mean free path of gas particles.

As usual, the linearized hydrodynamic equations for the perturbations $\left\{\delta n(\mathbf{r}; t), \delta \mathbf{U}(\mathbf{r}; t), \delta T(\mathbf{r}; t)\right\}$ are written in the Fourier space. A set of Fourier transformed dimensionless variables are then introduced as
\begin{equation}
\label{3.1}
\rho_{{\bf k}}(\tau)=\frac{\delta n_{{\bf k}}(\tau)}{n_{\text{H}}}, \
{\bf w}_{{\bf k}}(\tau)=\frac{\delta {\bf U}_{{\bf
k}}(\tau)}{v_{0\text{H}}},\
\theta_{{\bf k}}(\tau)=\frac{\delta
T_{{\bf k}}(\tau)}{T_{H}},
\end{equation}
where $\delta y_{{\bf k}\alpha}\equiv \{\delta n_{{\bf k}},{\bf w}_{{\bf k}}(\tau), \theta_{{\bf k}}(\tau)\}$ is defined as
\begin{equation}
\label{3.2}
\delta y_{{\bf k}\alpha}(\tau)=\int d{\boldsymbol {\ell}}\;
e^{-i {\bf k}\cdot {\boldsymbol {\ell}}}\delta y_{\alpha}
({\boldsymbol {\ell}},\tau).
\end{equation}
Note that in Eq.\ (\ref{3.2}) the wave vector ${\bf k}$ is dimensionless.

After some straightforward algebra, linearization of the NS equations in $\rho$, $\mathbf{w}$, and $\theta$ shows that the $d-1$ transverse velocity components ${\bf w}_{{\bf k}\perp}={\bf w}_{{\bf k}}-({\bf w}_{{\bf k}}\cdot
\widehat{{\bf k}})\widehat{{\bf k}}$ (orthogonal to the wave vector ${\bf k}$)
decouple from the other three modes and hence can be obtained easily. They are
%given by
\beq
\label{3.3}
{\bf w}_{{\bf k}\perp}(\tau)={\bf w}_{{\bf k}\perp}(0) \text{e}^{-\frac{1}{2}\eta^* k^2 \tau},
\eeq
where we have taken into account that $\eta^*$ does not depend on time in the HSS. Thus, since $\eta^*>0$, then the $d-1$ transversal shear modes ${\bf w}_{{\bf k}\perp}( \tau)$ are linearly stable.

%\begin{widetext}
The remaining (longitudinal) modes correspond to $\rho_{{\bf k}}$, $\theta_{{\bf k}}$, and
the longitudinal velocity component of the velocity field, $w_{{\bf k}||}={\bf w}_{{\bf
k}}\cdot \widehat{{\bf k}}$ (parallel to ${\bf k}$). These modes are coupled and obey the equation
\begin{equation}
\frac{\partial \delta y_{{\bf k}\alpha }(\tau )}{\partial \tau }=M_{\alpha \beta}
 \delta y_{{\bf k}\beta }(\tau ),
\label{3.4}
\end{equation}
where $\delta y_{{\bf k}\alpha }(\tau )$ denotes now the set  $\left\{ \rho _{{\bf k}},\theta _{{\bf k}},
 w_{{\bf k}||}\right\}$ and $\mathsf{M}$ is the square matrix
\begin{equation}
%\mathsf{M}=
\left(
\begin{array}{ccc}
0 & 0 & -i k \\
-\frac{d+2}{2(d-1)}\mu^*k^2&-2\bar{\zeta}_0-\frac{d+2}{2(d-1)}\kappa^*k^2 & -i k\left(\frac{2}{d}p^*+\zeta_U\right)\\
-i k p^* C_\rho & -i k \left(p^*+\Psi_p\right) &-\frac{d-1}{d}\eta^*k^2-\frac{1}{2}\gamma^*k^2
\end{array}
\right).
\label{3.5}
\end{equation}
%\end{widetext}
Here, $g(\phi)=1+\phi \partial_\phi \ln \chi(\phi)$, $C_\rho(\phi)=1+g(1-p^{*-1})$, and it is understood that $p^*$, $\eta^*$, $\gamma^*$, $\kappa^*$, $\mu^*$, and $\zeta_U$ are evaluated in the HSS. In addition, for a two-dimensional system,
\beq
\label{3.5.0.2}
\bar{\zeta}_0\equiv T_\text{H}\Bigg(\frac{\partial \zeta_0^*}{\partial T}\Bigg)_\text{s}=\chi \Delta_\text{s}^*\Bigg(\frac{1}{2}\sqrt{\frac{\pi}{2}}\al+\Delta_\text{s}^{*}\Bigg),
\eeq
\beq
\label{3.5.0.1}
\Psi_p\equiv T_\text{H}\Bigg(\frac{\partial p^*}{\partial T}\Bigg)_\text{s}=-\frac{\sqrt{2}}{\pi}\phi \chi \Delta_\text{s}^*.
\eeq
The longitudinal three modes have the form $\exp[s_n(k) \tau]$ for $n=1,2,3$, where $s_n(k)$ are the eigenvalues of the matrix $\mathsf{M}$, namely, they are the solutions of the cubic equation
\beq
\label{3.5.0}
\begin{split}
 &\!\!\!s^3+\Bigg[2\overline{\zeta}_0+k^2\Big(\frac{\eta^*+\gamma^*}{2}+2\kappa^*\Big)\Bigg]s^2
+k^2\Bigg[k^2 \kappa^*
\left(\eta^*+\gamma^*\right)
\\
%\nonumber\\
%XXX \times \left(\eta^*+\gamma^*\right)XXX
&\!\!\!+p^*\Big(C_\rho+p^*+\zeta_U\Big)+\Psi_p\Big(p^*+\zeta_U\Big)+\Big(\eta^*+\gamma^*\Big)\overline{\zeta}_0\Bigg]s\\
&\!\!\!+2k^2\Bigg\{p^*C_\rho \overline{\zeta}_0+k^2\Big[p^*C_\rho \kappa^*-\Big(p^*+\Psi_p\Big)\mu^*\Big]\Bigg\}=0.
\end{split}
\eeq
For given values of $\alpha$ and $\phi$, the analysis of Eq.\ \eqref{3.5.0} shows that for $k\neq 0$ one mode is real while the other two are a complex conjugate pair of propagating modes.

For small $k$, the solution to Eq.\ \eqref{3.5.0} can be written as a perturbation expansion:
\beq
\label{n1}
s_{n}(k)=s_{n}^{(0)}+k s_n^{(1)}+k^2s_{n}^{(2)}+\cdots
\eeq
Substitution of the expansion \eqref{n1} into the cubic equation \eqref{3.5.0} yields $s_1^{(0)}=s_2^{(0)}=0$, $s_3^{(0)}=-2\overline{\zeta}_0$,
%\beq
%\label{n2}
%s_1^{(0)}=s_2^{(0)}=0, \quad s_3^{(0)}=-2\overline{\zeta}_0,
%\eeq
\begin{eqnarray}
%\beq
%\label{n3}
&& s_1^{(1)}=-s_2^{(1)}=i \sqrt{p^* C_\rho}, \quad s_3^{(1)}=0,\nonumber\\
%\eeq
%
%\label{3.5.1}
&&s_{1}^{(2)}=s_{2}^{(2)}=-\frac{X-p^*C_\rho}{4\bar{\zeta}_0}, \quad \nonumber\\
% s_{3}^{(2)}=-\frac{p^* C_\rho-X+2Y\overline{\zeta}_0}{2\overline{\zeta}_0}.
&&s_{3}^{(2)}=\frac{(p^*+\zeta_U)(p^*+\Psi_p)-4\overline{\zeta}_0\kappa^*}{2\overline{\zeta}_0}.
\end{eqnarray}
Here,
\beq
\label{3.5.2}
X=p^*\Big(C_\rho+p^*+\zeta_U\Big)+\Psi_p\Big(p^*+\zeta_U\Big)+\Big(\eta^*+\gamma^*\Big)\overline{\zeta}_0.
\eeq
% \beq
% \label{3.5.3}
% Y=\frac{\eta^*+\gamma^*}{2}+2\kappa^*.
% \eeq
Since the Navier-Stokes hydrodynamic equations  are valid to second order in $k$, the above perturbation solutions are relevant to the same order. In particular, in the extreme long wavelength limit ($k=0$, inviscid fluid or Euler hydrodynamic order), two of the eigenvalues are zero (marginally stable solution) and the third one is negative (stable solution).

\begin{figure}[h]
\centering
\includegraphics[width=0.35 \textwidth,clip]{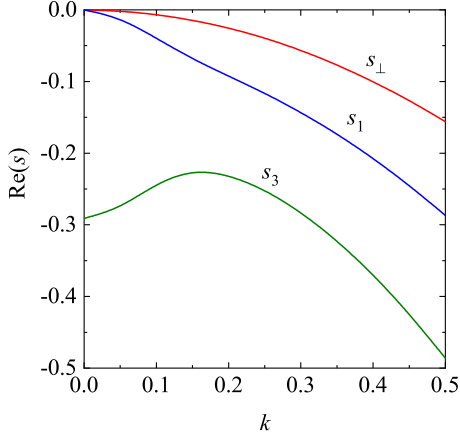}
\caption{Dispersion relations for a granular fluid with $\alpha=0.8$ and
$\phi=0.2$. From top to bottom the curves correspond to the $d-1$ degenerate shear (transversal) modes and
the remaining longitudinal modes. Only the real parts of the eigenvalues is plotted.}
\label{fig2}       % Give a unique label
\end{figure}

An analysis of the eigenvalues of the matrix $\mathsf{M}$ for finite $k$ shows that in general $\text{Re}(s_n)\leq 0$ and hence the HSS is linearly \emph{stable} in the complete range of values of the wave number $k$ studied. As an illustration, the dispersion relations $s_{n}(k)$ for a fluid with $\alpha=0.8$ and $\phi=0.2$ are plotted in Fig.\ \ref{fig2}. Only the real part (propagating modes) of eigenvalues is represented. For $d=2$, $\chi(\phi)=(1-\frac{7}{16}\phi)/(1-\phi)^2$ \cite{JM87}.

%\section{Conclusions}

In summary, a linear stability analysis of the HSS of a confined granular system has been performed in the context of the inelastic Enskog equation. Our study (i) takes into account the nonlinear dependence of the NS transport coefficients and the cooling rate on the coefficient of restitution and (ii) considers moderate densities. Thus, the present contribution covers some of the limitations of previous works \cite{BRS13,BBGM16}. Our results show no new surprises with respect to earlier works \cite{BRS13,BBGM16} since the HSS is linearly stable at finite dissipation and/or moderate density. This conclusion contrasts with the one found in the conventional IHS model where it was shown that the resulting hydrodynamic equations exhibit a long wavelength instability for $d-2$ of the hydrodynamic modes \cite{G05}.

%\section{Acknowledgments}

The research of V.G.~has been supported by the Spanish Ministerio de Econom{\'i}a y Competitividad through Grant No.~FIS2016-76359-P and by the Junta de Extremadura (Spain) Grant Nos.~IB16013 and GR18079, partially financed by ``Fondo Europeo de Desarrollo Regional'' funds. The work of R.B.~has been supported by the Spanish Ministerio de Econom{\'i}a y Competitividad through Grant No.~FIS2017-83709-R. The research of R.S.~has
been supported by the Fondecyt Grant No.~1180791 of ANID (Chile).


\begin{thebibliography} {99}
\bibitem{note}As shown in Ref.\ \cite{BBMG15}, there is a nonzero contribution to the coefficient $\zeta_U$ for dilute granular gases when the kurtosis $a_{2,s}$ (which measures the departure of the HSS from the Maxwellian distribution) is accounted for. However, the magnitude of $\zeta_U$ is very small for $\phi=0$.
\end{thebibliography}
\end{document}